\documentclass[twocolumn,secnumarabic,amssymb, amsmath, nofootinbib,tightenlines,
nobibnotes, aps, prl,epsfig]{revtex4}
\usepackage{graphicx}% Include figure files
\usepackage{dcolumn}% Align table columns on decimal point
\usepackage{bm}% bold math
\begin{document}
\preprint{APS/123-QED}
\title{Collinear approach for top quark production at  ep colliders}% Force line breaks with \\

\author{G.R.Boroun}%
 \email{grboroun@gmail.com; boroun@razi.ac.ir }
\affiliation{  Department of physics, Razi University, Kermanshah
67149, Iran}% \textbackslash\textbackslash
\date{\today}% It is always \today, today,
\begin{abstract}
%%%%%%%%%%%%%%%%%%%%%%%%%%%%%%%%%%%%%%%%%%%%%%%%%%%%%%%
An approximately analysis study of the collinear approach for
$t\overline{t}$ production at the Future Circular Collider
hadron-electron (FCC-he) and the Large Hadron electron Collider
(LHeC) is performed. To study heavy quark production processes,
the collinear generalized double asymptotic scaling (DAS) approach
in a wide rang of $Q^{2}$ is used. The reduced cross sections for
top quark pair production at the FCC-he (for
$Q^{2}{\geq}m^{2}_{t}$) and LHeC (for $Q^{2}{\leq}m^{2}_{t}$)
center-of-mass energies at the high inelasticity are derived.
These results may be considering for both the future collider
experiments at center-of-mass energy frontier and the improvement
of the phenomenological models on kinematics of low values of the
Bjorken variable $x$. The evolution of the gluon density for
heavy-quark production in DIS is obtained at low $x$ to include
effects of heavy-quark masses. A nonlinear modification of these
results shown that the nonlinear corrections, with respect to
the top quark mass, are very small for $Q^{2}{\propto}m^{2}_{t}$.\\

%%%%%%%%%%%%%%%%%%%%%%%%%%%%%%%%%%%%%%%%%%%%%%%%%%%%%%%
\end{abstract}
 \pacs{***}%PACS, the Physics and Astronomy
                              %Classification Scheme.
 %Use showkeys class option if keyword
\maketitle
%\tableofcontents
%**********************************************************
\subsection{I. Introduction}

The Large Hadron electron Collider (LHeC) [1] and the Future
Circular Collider hadron-electron (FCC-he) [2] are  proposed
facilities of using newly built electron-proton center of mass
energies at $\sqrt{s}~{\cong}~ 1.3$ and $3.5~\mathrm{TeV}$,
respectively. At the LHeC and FCC-he, it is foreseen that a
$60~\mathrm{GeV}$ electron beam collides with one of the
$7~\mathrm{TeV}$ intense proton beams of the LHC, or with a
$50~\mathrm{TeV}$ proton beam of a Future Circular Collider for
hadron-hadron scattering (FCC-hh). The LHeC and FCC-he
measurements can be performed with much increased precision and
extended to much lower values of $x$ and high $Q^{2}$. The LHeC
provides data on the charm and bottom structure functions
extending the range in $x,Q^{2}$ by up to 5 or 6 orders of
magnitude, respectively, compared to HERA. These new colliders
will be devoted to probe the energy frontier and complement the
discovery potential of the LHC with measurements of deep inelastic
scattering (DIS). The LHeC and FCC-he collisions lead into the
region of high parton densities at low $x$ since the kinematic
reach goes up to $Q^{2}{\simeq}1~\mathrm{TeV^{2}}$ and
$x{\simeq}10^{-5}..10^{-6}$ for LHeC and $10^{-7}$ for FCC-he.
These kinematics are pertinent in investigations of lepton-hadron
processes at ultra-high energy (UHE) neutrino astroparticle
physics [3-4]. Moreover a similar very high energy
electron-proton/ion collider (VHEep) [5] has been suggested based
on plasma wakefield acceleration, albeit with very low luminosity.
The center-of-mass energy, in this collider, is close to
$10~\mathrm{TeV}$ which is relevant in investigations of new
strong interaction dynamics related to high-energy cosmic rays and
gravitational physics. It can be used to study the top-quark
physics [6-13].\\
In particular the LHeC and FCC-he will be provided a cleaner
environment for the study of top quark pair production. The top
quark will be produced in these future colliders in pairs
($t\overline{t}$) through quantum chromodynamics (QCD) processes,
mostly photon-gluon fusion (PGF) at ultra-high energies
$\gamma^{*}+g{\rightarrow}t+\overline{t}$. The top quark pair
production cross sections are $\sigma_{t\overline{t}}=984.5~
\mathrm{pb}$ at $\sqrt{s}=14~\mathrm{TeV}$ and
$\sigma_{t\overline{t}}=826.4~ \mathrm{pb}$ at
$\sqrt{s}=13~\mathrm{TeV}$ at CMS [18] and ATLAS [19]
respectively. The top quark at the LHC is thereby characterized by
final states comprising the decay products of the two $W$ bosons
and two $b$ jets, and  produced in nuclear collisions with respect
to the partonic subprocesses
\begin{eqnarray}
q+\overline{q}{\rightarrow}t+\overline{t}+X,\nonumber\\
g+g{\rightarrow}t+\overline{t}+X,\nonumber\\
q+g{\rightarrow}t+\overline{t}+X,\nonumber
\end{eqnarray}
where $\sigma^{t\overline{t}}_{\mathrm{tot}}$ can be written as
\begin{eqnarray}
\sigma^{t\overline{t}}_{\mathrm{tot}}=\sum_{ij}\int_{0}^{4m_{t}^{2}/s}{d\beta}
\Phi_{ij}(\beta,\mu_{F}^{2})\widehat{\sigma}_{ij}(\beta,m_{t},\mu_{F}^{2},\mu_{R}^{2}).
\end{eqnarray}
Here $\Phi_{ij}$ is the partonic flux which is a convolution of
the densities of partons $i, j$ and $\widehat{\sigma}_{ij}$ is the
total partonic cross section for the inclusive production of a
heavy quark from partons $i, j$ [14-17]. The dimensionless
variable $\beta^{2}=1-\rho$, with $\rho=4m^{2}_{t}/s$, is the
squared relative velocity of the final state top quarks having
pole mass $m_{t}$ and produced at the square of the gluonic center
of mass energy $s$. The gluon-gluon fusion (GGF),
$g+g{\rightarrow}t+\overline{t}$, is dominantly at ultra-high
energies for $\mu=m_{t}$, thus partonic cross section is
summarized as follows
\begin{eqnarray}
\widehat{\sigma}_{gg}(\beta,m_{t})=\frac{\alpha_{s}^{2}}{m^{2}_{t}}
\{ f_{gg}^{(0)}+\alpha_{s}f_{gg}^{(1)}+\alpha_{s}^{2}f_{gg}^{(2)}+
\mathcal{O}{ \alpha_{s}^{3}} \}.
\end{eqnarray}
The functions $f_{gg}$ are known at leading and high-order
approximations and depend only on dimensionless parameters $\beta$
and $\rho$ [15].\\
The total cross section for the single top quark production via
charged current (CC) DIS scattering at the LHeC is
$1.89~\mathrm{pb}$ [1] due to the center-of-mass energy of
$1.3~\mathrm{TeV}$ and at the FCC-he is $15.3~\mathrm{pb}$ [2] due
to the center-of-mass energy of $3.5~\mathrm{TeV}$. It is
$0.05~\mathrm{pb}$ [1] for top quark pair production in
$t\overline{t}$ photoproduction mode at the LHeC and
$1.14~\mathrm{pb}$  at the FCC-he [2]. It means that the top quark
with mass about $172.76{\pm}0.3~\mathrm{GeV}$ [18] can be studied
at future colliders. It derives its mass from its coupling to the
Higgs Boson [19]. The $t\overline{t}$ productions at the LHC are
implemented  and used to extract PDFs [20]. In the future ep
collider,  top quark pair production via NC ep scattering will be
considered. Such future ep collider facilities allow
high-precision measurements of certain top quark properties
and have high sensitivity in finding new physics beyond the SM.\\
In this work, in the kinematic regime of the FCC-he, the top quark
pair production cross section is calculated in the collinear
approach. In sec. II, the theoretical framework for the
coefficient function in the collinear approach is presented. In
sec. III, the numerical results for  the reduced cross section
$\sigma^{t\overline{t}}_{r}(x,Q^{2})$ and the ratio
$R^{t\overline{t}}=F_{L}^{t\overline{t}}/F_{2}^{t\overline{t}}$,
in the kinematic regime of the FCC-he,
 are presented.\\

\subsection{II. Theory}

 The heavy quark pair production at HERA is obtained by fixed number
 of parton densities (fixed flavour number schemes, FFNS) close to threshold $\mu^{2}\sim m_{\mathcal{Q}}^{2}$ (where $m_{\mathcal{Q}}$ is the heavy quark mass).
  The
resummation of collinear logarithms
$\ln(\mu^{2}/m_{\mathcal{Q}}^{2})$ at scales far above the
threshold $\mu^{2}{\gg}m_{\mathcal{Q}}^{2}$ is achieved through
the use of variable flavour number schemes (VFNS). When the scale
is increased above heavy quark mass thresholds, the number of
active flavors increases in VFNS. For realistic kinematics it has
to be extended to the case of a general-mass VFNS (GM-VFNS) which
is defined similarly to the zero-mass VFNS (ZM-VFNS) in the
$Q^{2}/m_{\mathcal{Q}}^{2}{\rightarrow}\infty$ limit [21,22]. In
GM-VFNS the transition, from $n_{f}$ active flavors to $n_{f}+1$,
is considered in the construction of the charm-quark parton
distribution function. At some rather large scales (i.e.,
$Q^{2}>m^{2}_{\mathcal{Q}}$) the transition to two massive quarks
(i.e., $n_{f}{\rightarrow}n_{f}+2$) has been discussed in
Refs.[23,24]. In the GM-VFNS at high $Q^{2}$, the heavy-flavor
structure functions depend on the active flavor number since here
$n_{f}=4$ for $m_{c}^{2}<\mu^{2}<m_{b}^{2}$, $n_{f}=5$ for
$m_{b}^{2}<\mu^{2}<m_{t}^{2}$ and $n_{f}=6$ for
$\mu^{2}{\geq}m_{t}^{2}$ is chosen.\\
Recently in Ref.[25] authors studied the transverse momentum
dependent (TMD) gluon distribution function in heavy quark
production processes. They used the Kimber-Martin-Ryskin [26]
prescription from the Bessel inspired behavior of parton densities
at small $x$. The heavy quark reduced cross section is defined in
terms of the heavy quark structure functions as
\begin{eqnarray}
\sigma_{r}^{\mathcal{Q}\overline{\mathcal{Q}}}(x,Q^{2})=F_{2}^{\mathcal{Q}\overline{\mathcal{Q}}}(x,Q^{2})
-f(y)F_{L}^{\mathcal{Q}\overline{\mathcal{Q}}}(x,Q^{2}),
\end{eqnarray}
where $f(y)=y^{2}/1+(1-y)^{2}$ and $y$ is the inelasticity. In the
collinear generalized double asymptotic scaling (DAS) [27]
approach, the heavy quark structure functions are driven by the
following form
\begin{eqnarray}
F_{k=2,
L}^{{\mathcal{Q}}\overline{{\mathcal{Q}}}}(x,Q^{2})=\sum_{i=g,q,\overline{q}}
C_{k,i}^{{\mathcal{Q}}\overline{{\mathcal{Q}}}}(x, \mu^{2},
m^{2}_{\mathcal{Q}} ){\otimes} xf_{i}(x,\mu^{2}),
\end{eqnarray}
where $xf_{i}$ are the parton distribution functions (PDFs) and
$C_{k,i}^{{\mathcal{Q}}\overline{{\mathcal{Q}}}}(x, Q^{2})$ are
the DIS coefficient functions. The gluon density is dominant for
$x<0.1$, therefore a further simplification with the contribution
due to PGF is obtained by neglecting the contributions due to
incoming light quarks and antiquarks in Eq. (3), which is
justified because they vanish at LO and are numerically suppressed
at high order corrections as
\begin{eqnarray}
F_{k=2, L}^{{Q}\overline{{Q}}}(x,Q^{2})\simeq
C_{k,g}^{{Q}\overline{{Q}}}(x, \mu^{2}, m^{2}_{Q} ){\otimes}
xf_{g}(x,\mu^{2}),
\end{eqnarray}
where $xf_{g}(x,Q^{2})=G(x,Q^{2})$ is the gluon distribution
function and the $\otimes$ symbol denotes the convolution integral
which turns into a simple multiplication in Mellin $N$-space. The
notation is defined by $a(x)\otimes
b(x)=\int_{x}^{1}\frac{dz}{z}a(z)b(\frac{x}{z})$. The coefficient
functions at leading-order  up to next-to-next-to leading order
(NNLO) approximations [25] read
\begin{eqnarray}
C^{\mathrm{LO}}_{k,g}(x,\mu^{2})&=&e_{Q}^{2}a_{s}(\mu^{2})B_{k,g}^{(0)}(x,a),\nonumber\\
C^{\mathrm{NLO}}_{k,g}(x,\mu^{2})&=&e_{Q}^{2}a_{s}(\mu^{2})[B_{k,g}^{(0)}(x,a)
+a_{s}(\mu^{2})B_{k,g}^{(1)}(x,a)],\nonumber\\
C^{\mathrm{NNLO}}_{k,g}(x,\mu^{2})&=&e_{Q}^{2}a_{s}(\mu^{2})[B_{k,g}^{(0)}(x,a)
+a_{s}(\mu^{2})B_{k,g}^{(1)}(x,a)\nonumber\\
&&+a^{2}_{s}(\mu^{2})B_{k,g}^{(2)}(x,a)],
\end{eqnarray}
where $a=m^{2}/Q^{2}$ and
$a_{s}(\mu^{2})=\alpha_{s}(\mu^{2})/4\pi$. The explicit expression
for the coefficient function is relegated in Appendix A. The
default renormalisation  and factorization scales are set to be
equal $\mu_{R}^{2}=Q^{2}+4m^2$ and $\mu_{F}^{2}=Q^{2}$. In order
to fix the unphysical mass scale $\mu$, the renormalisation and
factorisation scale for the heavy quarks is set to
$\mu^{2}=Q^{2}+4m^2$. Thereofe, the top quark structure functions
at low $x$ are given by the following form
\begin{eqnarray}
F_{k}^{{t}\overline{{t}}}(x,Q^{2})=C_{k,g}^{{t}\overline{{t}}}(x,
\mu^{2}, m^{2}_{t} ){\otimes} G_{n_{f}=6}(x,\mu^{2}),
\end{eqnarray}
where $G_{n_{f}}$ is the gluon distribution function due to the
number of active quark flavors.\\
In the presence of heavy quarks, the gluon distribution function
$G_{n_{f}}(x,\mu^{2})$ is defined  with respect to threshold of
heavy quark pair production. In Ref.[28], the authors used  a
simplified version of the method introduced by Aivazis, Collins,
Olness, and Tung (ACOT) [29] using a rescaling variable
$x_{i}=x\eta_{i}(Q^{2})$ where
$i{\in}~u,\overline{u},d,\overline{d},s,\overline{s},c,\overline{c},b,\overline{b}$
and $\eta_{i}(Q^{2})=1+4M^{2}_{i}/Q^{2}$ . The authors of [28]
showed that the massless gluon distribution treating active flavor
effects properly is defined as
\begin{eqnarray}
G_{nf=4}(x,Q^{2})& =& \frac{3}{5}G_{nf=3}(x,Q^{2}),
\end{eqnarray}
and
\begin{eqnarray}
G_{nf=5}(x,Q^{2})& =& \frac{6}{11}G_{nf=3}(x,Q^{2}).
\end{eqnarray}
$G_{nf=3}(x,Q^{2})$ is dependent on $F_{2}(x,Q^{2})$ and
$\partial{F_{2}(x,Q^{2})}/\partial{\mathrm{\ln}}x$ using a Laplace
transformation method. Indeed authors in Ref.[28] have
demonstrated that a parametrization of the ZEUS experimental data
on the proton structure function $F_{2}(x,Q^{2})$ gives an
analytic solution  for the LO gluon distribution
$G_{nf=3}(x,Q^{2})$ as a function of $x$ and $Q^{2}$. This
function (i.e., $G_{nf=3}$ ) is defined very well for low $x$
($0<x~{\leq}~0.06$) by an expression quadratic in both
$\ln(Q^{2})$ and $\ln(1/x)$ as
\begin{eqnarray}
G_{nf=3}(x,Q^{2})=-2.94-0.359{\ln}(Q^{2})-0.101{\ln^{2}}(Q^{2})\nonumber\\
+(0.594-0.0792{\ln}(Q^{2})-0.000578{\ln^{2}}(Q^{2})){\ln}(1/x)\nonumber\\
+(0.168+0.138{\ln}(Q^{2})+0.0169{\ln^{2}}(Q^{2})){\ln}(1/x)
\end{eqnarray}
The gluon distribution function $G_{nf}(x,Q^{2})$ is determined by
the measured $F_{2}(x,Q^{2})$ for deep inelastic $\gamma^{*}p$
scattering to include the effects of heavy-quark masses. Several
methods for the determination of the charm and bottom quark
structure functions in the
nucleon have been proposed in Refs.[30-33].\\
The evolution equation for the proton structure function at LO
approximation reads
\begin{eqnarray}
\frac{1}{x}\frac{\partial{F_{2}(x,Q^{2})}}{\partial{{\ln}Q^{2}}}=\frac{\alpha_{s}}{4\pi}
[\int_{x}^{1}\frac{dz}{z^{2}}F_{2}(z,Q^{2})K_{qq}(\frac{x}{z})\nonumber\\
+\sum_{i}e_{i}^{2}\frac{1}{\eta_{i}}\int_{x}^{1}\frac{dz}{z^{2}}
G(\eta_{i}z,Q^{2})K_{gq}(\frac{x}{z})],
\end{eqnarray}
where $e_{i}^{2}$ is the squares of the quark charges and $K^{,}$s
are the splitting functions. In Eq.(11) the gluon distribution
function shift from $z$ to $\eta_{i}z$ for activation of heavy
quarks. When taking mass effects into account, the exact solution
for the gluon distribution at $n_{f}=5$ has the following form
\begin{eqnarray}
G(x,Q^{2})&=&G_{n_{f}=3}(x,Q^{2})+\sum_{n=1}^{N}(-1)^{n}\sum_{k=0}^{n}\left(
{\begin{array}{c}
n  \\
   k  \\
  \end{array} } \right)\nonumber\\
&&\alpha^{n-k}\beta^{k}G_{n_{f}=3}(\eta_{c}^{n-k}\eta_{b}^{k}x,Q^{2}).
\end{eqnarray}
Here $\eta_{c}=1+4\frac{M_{c}^{2}}{Q^{2}}$,
$\eta_{b}=1+4\frac{M_{b}^{2}}{Q^{2}}$, $\alpha=(2/3\eta_{c})$,
$\beta=(1/6\eta_{b})$ and $\big( ^{n} _{k} \big)$ is a binomial
coefficient. The summations in Eq.(12) are finite as the sum on
$n$ terminates at $N$ such that $(N+1)\ln{\eta_{c}}{\geq}\ln(1/x)$
[28]. Indeed the ratio of the gluon distributions
$\frac{G_{n_{f}+j}(x,Q^{2})}{G_{n_{f}}(x,Q^{2})}$ ,for $j=1,2,3$
and $n_{f}=3$, are finite by the following form
\begin{eqnarray}
\frac{G_{n_{f}+j}(x,Q^{2})}{G_{n_{f}}(x,Q^{2})}=\frac{
\sum_{i}^{n_{f}}e^{2}_{i}}{\sum_{i}^{n_{f}+j}e^{2}_{i}}
\end{eqnarray}
where for $j=1,2$ and $3$ the ratio is $3/5, 6/11$ and $2/5$,
respectively. Thus, the gluon distribution function for $n_{f}=6$
at $Q^{2}{\geq}m^{2}_{t}$ is obtained into the gluon distribution
function of three massless quarks,$G_{n_{f}=3}$, by the following
form
\begin{eqnarray}
G_{n_{f}=6}(x,Q^{2})=\frac{2}{5}G_{n_{f}=3}(x,Q^{2}).
\end{eqnarray}
This behavior of the gluon distribution function for massive c,b
and t quarks [28, 34] is shown in Fig.1 over a wide range of
$Q^{2}$ for $x=10^{-3}$. In this figure (i.e., Fig.1) we display
the gluon distribution function vs. $Q^{2}$ for $n_{f}=3, 4, 5,
6$. We observe that the  gluon distribution is reduced as a
function of $n_{f}$ [34]. In the following we use the massless
gluon distribution functions (i.e., Eqs.(10), (8), (9) and (14))
for $n_{f}=3, 4, 5, 6$ respectively as compared together in Fig.1
as the solid- black, dash-red, dot-green and dash-dot-blue
curves.\\
\begin{figure} \centering
\includegraphics[width=0.45\textwidth]{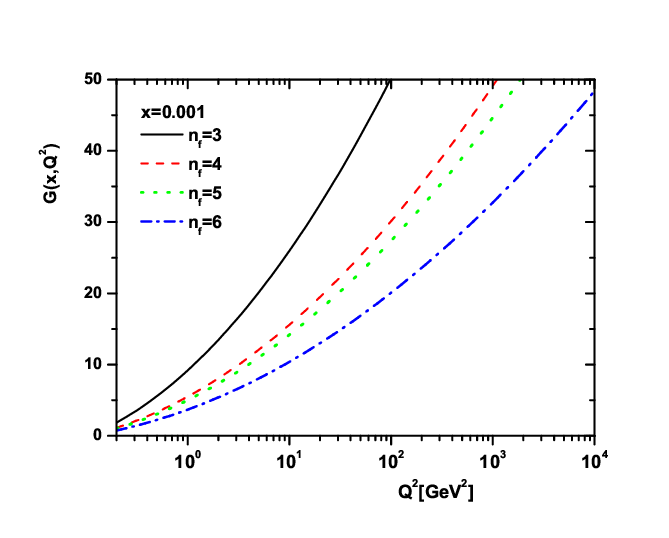}
\caption{ The gluon distribution function at $x=0.001$ vs.$Q^{2}$
for $n_{f}=\{3,4,5,6\}$. $n_{f}=3$ is the largest and $n_{f}=6$ is
the smallest curve.} \label{Fig1}
\end{figure}
Therefore, the top reduced cross section is defined as
\begin{eqnarray}
\sigma_{r}^{t\overline{t}}(x,Q^{2})&=&[C_{2,g}^{{t}\overline{{t}}}(x,
Q^{2}, m^{2}_{t} ) -f(y)C_{L,g}^{{t}\overline{{t}}}(x, Q^{2},
m^{2}_{t} )]\nonumber\\
&&{\otimes} G_{n_{f}=6}(x,Q^{2}).
\end{eqnarray}
With the explicit form of the coefficient functions, the top
reduced cross section $\sigma_{r}^{t\overline{t}}(x,Q^{2})$ is
extracted from the parametrization of $G_{n_{f}}(x,Q^{2})$ for
$Q^{2}{\geq}m^{2}_{t}$ in the collinear
approach.\\
In order to make sure these results are valid in the deep
inelastic region,  the nonlinear  corrections to the gluon
distribution is taken into account. It is known that the
absorptive corrections (or gluon recombination effects) in the low
$x$, low $Q^{2}$ region are not negligible and reduce the growth
of the gluon parton distribution function. Indeed the fusion
processes, $gg{\rightarrow}g$, for small momentum transfer in the
transverse area become important [35-41]. The theoretical
predictions of these effects were first emphasized
long ago by Gribov, Levin and Ryskin [42] and followed by Mueller and Qiu (MQ) [43].\\
The evolution of the gluon density is modified at low values of
$x$ by an extra nonlinear term, quadratic in the gluon density, as
\begin{eqnarray}
\frac{\partial{G(x,Q^{2})}}{\partial{\ln}Q^{2}}&=&\frac{\partial{G(x,Q^{2})}}{\partial{\ln}Q^{2}}|_{DGLAP}\nonumber\\
&&-\frac{81}{16}\frac{\alpha_{s}^{2}(Q^{2})}{\mathcal{R}^{2}Q^{2}}\int_{\chi}^{1}\frac{dz}{z}G^{2}(\frac{x}{z},Q^{2}).
\end{eqnarray}
where $\chi=\frac{x}{x_{0}}$ and $x_{0}$ is the boundary condition
that at $x{\geq}x_{0}(=10^{-2})$ the nonlinear corrections are
negligible. The correlation length $\mathcal{R}$  determines the
size of the nonlinear terms, as $\mathcal{R}\sim 1$ is of the
order of the proton radius. The $\mathcal{R}$ is approximately
equal to $\simeq 5~\mathrm{GeV}^{-1}$ if the gluons are populated
across the proton and it is equal to $\simeq 2~\mathrm{GeV}^{-1}$
if the gluons have hot-spot like structure. Eq.(16) leads to
saturation of the gluon density at low $Q^{2}$ with decreasing $x$
as the nonlinear correction (NLC) to the gluon distribution is
defined by the following form [44]
\begin{eqnarray}
G^{\mathrm{NLC}}(x,Q^{2})&=&G^{\mathrm{NLC}}(x,Q_{0}^{2})+[G(x,Q^{2})-G(x,Q_{0}^{2})]\nonumber\\
&&-\int_{Q_{0}^{2}}^{Q^{2}}\frac{81}{16}\frac{\alpha_{s}^{2}(Q^{2})}{\mathcal{R}^{2}Q^{2}}\int_{\chi}^{1}\frac{dz}{z}G^{2}(\frac{x}{z},Q^{2})d{\ln}Q^{2}\nonumber\\
\end{eqnarray}
where $G(x,Q^{2})$ and $G(x,Q_{0}^{2})$ are the unshadowed gluon
distributions. At the initial scale $Q_{0}^{2}$, the low $x$
(i.e., $x<x_{0}$ ) behavior of the nonlinear gluon distribution is
assumed to be [45]
\begin{eqnarray}
G^{\mathrm{NLC}}(x,Q_{0}^{2})&=&G(x,Q_{0}^{2})\{1+\frac{27\pi{\alpha_{s}(Q_{0}^{2})}}{16\mathcal{R}^{2}Q_{0}^{2}}\theta(x_{0}-x)\nonumber\\
&&{\times}[G(x,Q_{0}^{2})-G(x_{0},Q_{0}^{2})] \}^{-1}.
\end{eqnarray}
In the following, the role of the nonlinear effects on the
behavior of the heavy quark distribution functions in the low $x$
region will be considered.\\
%%%%%%%%%%%%%%%%%%%%%%%%%%%%%%%%%%%%%%%%%%%%%%%%%%%%%%%%%%%%%%
\begin{figure} \centering
\includegraphics[width=0.45\textwidth]{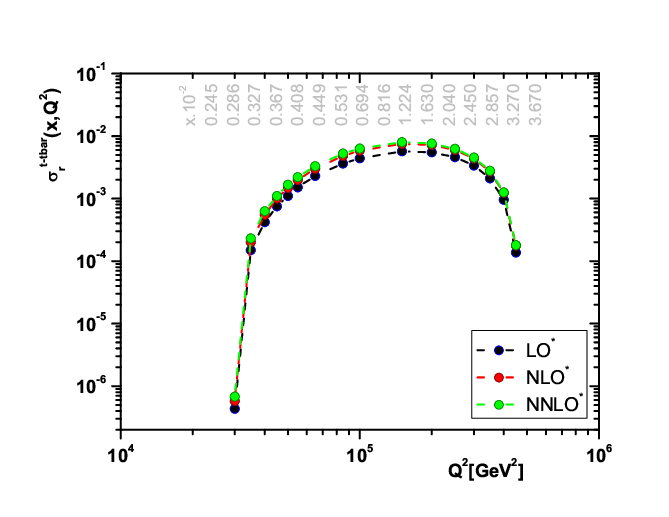}
\caption{The reduced top cross sections
$\sigma^{t\overline{t}}_{r}(x,Q^{2})$ plotted at different $Q^{2}$
values for $\sqrt{s}=3.5~\mathrm{TeV}$ at $Q^{2}{\geq}m^{2}_{t}$.
The gray numbers on the upper scale of the figure are the Bjorken
$x$ values for each $Q^{2}$ where lie in the interval
($0.001<x<0.1$). The predictions are obtained using an analytical
parameterization of the gluon density at LO and the coefficient
functions at $\mathrm{LO}^{*}$ up to $\mathrm{NNLO}^{*}$
approximations. These results represent the upper bound for top
reduced cross section from the high inelasticity (i.e., $y=1$).}
\label{Fig2}
\end{figure}

%%%%%%%%%%%%%%%%%%%%%%%%%%%%%%%%%%%%%%%%%%%%%%%%%%%%
\subsection{III. Results and Discussions}

 In this analysis, the value of $\alpha_{s}(M_{z})$ is set
 to the world average of $\alpha_{s}(M_{z})=0.118$, in line with
the recommended value from the particle data group (PDG) [46]. In
accordance with the values recommended by the Higgs Cross Section
Working Group [47], the charm-quark, bottom-quark and top-quark
pole masses are set as in the NNPDF default analysis to
$m_{c}=1.65~ \mathrm{GeV}$ and $m_{b}=4.78~ \mathrm{GeV}$ and
$m_{t}=172.5~\mathrm{GeV}$ [48], respectively.\\
In Fig.2, the behavior of the top-quark reduced cross section,
with respect to the number of active flavors $n_{f}=6$, for
$Q^{2}{\geq}m_{t}^{2}$ at center-of-mass energy
$\sqrt{s}=3.5~\mathrm{TeV}$ at the high inelasticity (i.e., $y=1$)
is presented at LO up to NNLO approximations\footnote {The
coefficient functions in our results are demonstrated in an
approximation order as $\mathrm{LO}{\rightarrow}\mathrm{LO}^{*}$,
$\mathrm{NLO}{\rightarrow}\mathrm{NLO}^{*}$ and
$\mathrm{NNLO}{\rightarrow}\mathrm{NNLO}^{*}$ for in order to
distinguish these calculations from the exact calculations in LO,
NLO and NNLO}. In this figure we employ the LO analytic gluon
distribution together with the LO up to NNLO DAS coefficient
functions (in an approximately approach) in a range of the
kinematical variables $x$ and $Q^{2}$, $x<0.1$ (in what follows
the value of $x={Q^{2}}/{sy}$ for each $Q^{2}$ is provided on the
upper scale of the figure) and
$m_{t}^{2}{\leq}Q^{2}{\leq}15m_{t}^{2}$. To show the contribution
and importance of the longitudinal structure function
$F_{L}^{t}(x,Q^{2})$, the top reduced cross sections at
high-inelasticity $y=1$ are derived, as
\begin{eqnarray}
\sigma_{r}^{t\overline{t}}(x,Q^{2})&=&F_{2}^{t\overline{t}}(x,Q^{2})-F_{L}^{t\overline{t}}(x,Q^{2}),\nonumber\\
&&=[C_{2,g}^{{t}\overline{{t}}}(x, Q^{2}, m^{2}_{t} )
-C_{L,g}^{{t}\overline{{t}}}(x, Q^{2},
m^{2}_{t} )]\nonumber\\
&&{\otimes} G_{n_{f}}(x,Q^{2}).
\end{eqnarray}
Notice that the large inelasticity is only for scattered electron
energies much smaller than the electron beam energy, where the
electromagnetic and hadronic backgrounds are important. Eq.(19) is
an upper bound on the top reduced cross section at low $x$ in a
wide range of $Q^{2}$. It is shown that this bound can be used to
constrain the range of applicability of the linear and nonlinear
corrections to the top reduced cross section at the FCC-he
collider. Figure 2 clearly demonstrates that the extraction
procedure provides correct behaviors of the top reduced cross
section in all three, LO, NLO and NNLO approximations. In Fig.2 we
observe that the maximum value for the top reduced cross section
is obtained at $Q^{2}{\cong}5m^{2}_{t}$. In Fig.3, the
$Q^{2}$-dependence, at low $x$ for high inelasticity, of the top
reduced cross section for $Q^{2}{<}m_{t}^{2}$ at center-of-mass
energy $\sqrt{s}=1.3~\mathrm{TeV}$, in the LO and NLO
approximations is presented. These results determined by using the
number of active flavors $n_{f}=5$  which the gluon distribution
corresponds to ${6}/{11}G_{nf=3}(x,Q^{2})$. In Fig.4 the top
reduced cross section plotted as function of the center-of-mass
energy ,$\sqrt{s}$, for the high inelasticity at the Bjorken $x$
values $x=0.003$ and $0.0003$. In this figure the center-of-mass
energy range for the EIC to VHEeP is defined to be
$100~\mathrm{GeV}<\sqrt{s}<50~\mathrm{TeV}$. The behavior of the
top reduced cross section dependence on the gluon distribution
functions with respect to the number of flavors $n_{f}$ as
$n_{f}=5$ in the region $Q^{2}<m^{2}_{t}$ and $n_{f}=6$ in the
region $Q^{2}{\geq}m^{2}_{t}$.
%%%%%%%%%%%%%%%%%%%%%%%%%%%%%%%%%%%%%%%%%%%%%%%%%%
\begin{figure} \centering
\includegraphics[width=0.45\textwidth]{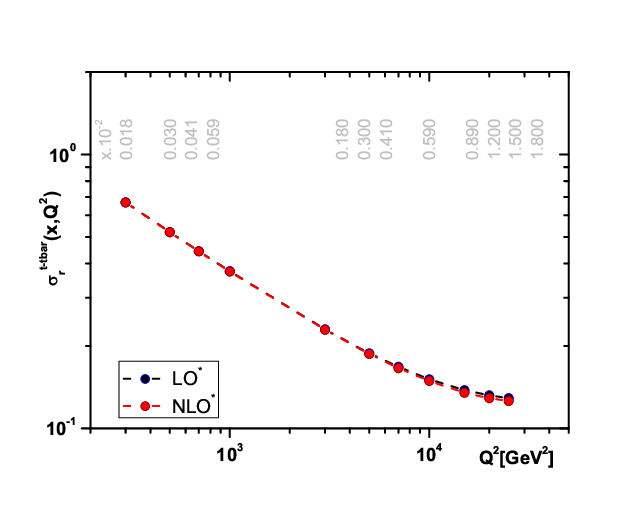}
\caption{ The same as Fig.2 for $\sqrt{s}=1.3~\mathrm{TeV}$ at
$Q^{2}{<}m^{2}_{t}$.} \label{Fig3}
\end{figure}
\begin{figure} \centering
\includegraphics[width=0.45\textwidth]{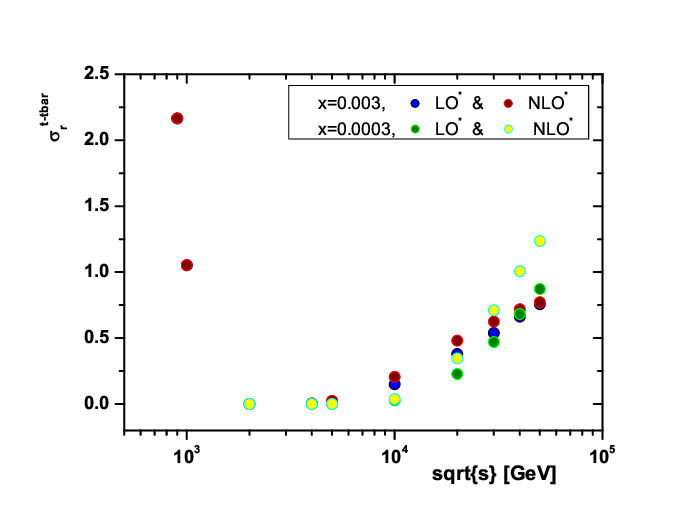}
\caption{The same as Fig.2 at fixed $x$ as a function of the
center-of-mass energy $\sqrt{s}$. The center-of-mass energy
corresponding to the chosen kinematics lies in the interval
$500~\mathrm{GeV}<\sqrt{s}\leq 50~\mathrm{TeV}$.} \label{Fig4}
\end{figure}
\begin{figure} \centering
\includegraphics[width=0.45\textwidth]{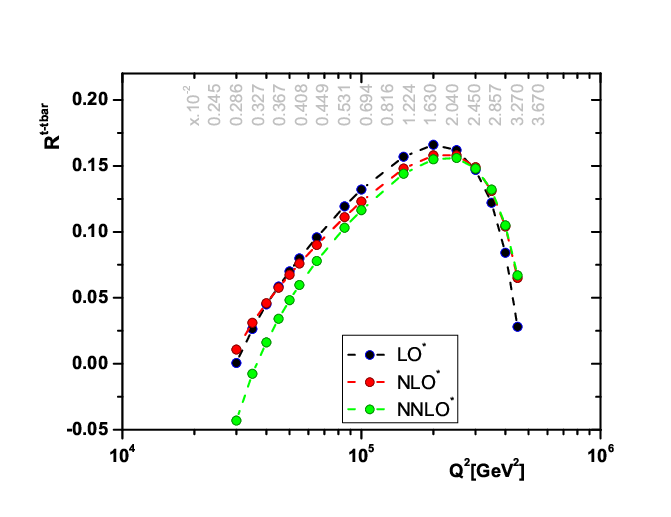}
\caption{The same as Fig.2 for the ratio
$R^{t\overline{t}}=\frac{F_{L}^{t\overline{t}}}{F_{2}^{t\overline{t}}}$
as a function of $Q^{2}$ for $\sqrt{s}=3.5~\mathrm{TeV}$.}
\label{Fig5}
\end{figure}
%%%%%%%%%%%%%%%%%%%%%%%%%%%%%%%%%%%%%%%%%%%%%%%%%%%%%
Figure 5 presents the ratio
$R^{t\overline{t}}=F_{L}^{t\overline{t}}/F_{2}^{t\overline{t}}$ as
a function of $Q^{2}$ in a wide range of $x$ at LO up to NNLO
approximations. This shows the importance of top quark
longitudinal structure function measurements [48] at energies of
future ep colliders. Since the difference between the estimated
$\sigma_{r}^{t\overline{t}}$ and $F_{2}^{t\overline{t}}$ is due to
the contribution of the longitudinal structure function
$F_{L}^{t\overline{t}}$, the calculations show that these
contributions are rather important at high $Q^{2}$ [49]. The
maximum value for the ratio $R^{t\overline{t}}$ is equal to
$\simeq 0.16$ for $Q^{2}{\simeq}5m_{t}^{2}$. For
$Q^{2}{\gtrsim}7m_{t}^{2}$, the ratio decreases as $Q^{2}$
increases. The maximum value of the ratio $R^{t\overline{t}}$ is
approximately constat [30] within the
center-of-mass energy region between HERA and the  FCC-he.\\
In Fig.6, the computed results of the nonlinear top reduced cross
section are compared with the linear behavior in the LO and NLO
approximations . This behavior is considered at  $x<10^{-2}$ for
$Q^{2}{\geq}m_{t}^{2}$ in the hot-spot point where the value of
this parameter is defined to be $R=2~\mathrm{GeV^{-1}}$ in this
paper. The nonlinear effects of the top reduced cross section are
very small due to the large top quark mass in a wide range of
$Q^{2}$. Consequently, knowledge of the linear and non-linear
corrections offers the possibility to perform cross-section
calculations of ultra-high-energy processes in heavy quark
production applying higher-order corrections.\\
\begin{figure} \centering
\includegraphics[width=0.45\textwidth]{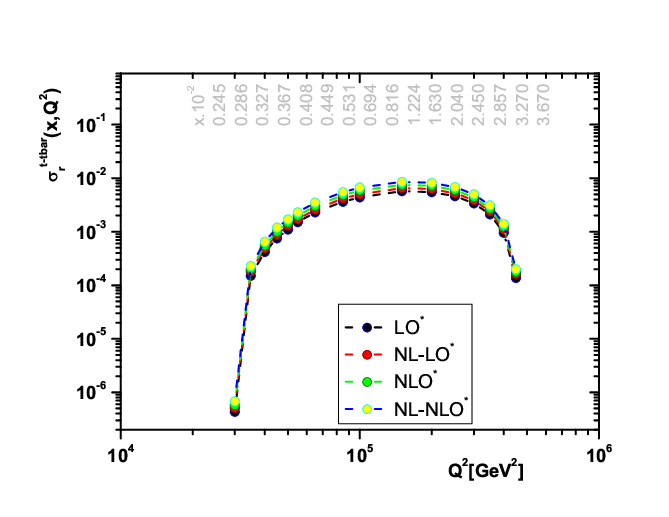}
\caption{The same as Fig.2 compared with the nonlinear corrections
to the reduced cross section in the $\mathrm{LO}^{*}$ and
$\mathrm{NLO}^{*}$ approximations at the hot-spot point(i.e.,
$R=2~\mathrm{GeV}^{-1}$ ). } \label{Fig6}
\end{figure}
To summarize, the collinear approach to obtain the heavy-quark
structure functions at small Bjorken $x$ values in the LO up to
NNLO approximations is studied. The method presents an application
of extraction of the gluon density due to the number of active
flavors. With respect to the parameterization of $F_{2}$, the
gluon density gives an excellent approximation behavior using a
simplified ACOT approximation if the heavy-quarks are properly
treated as massive. The nonlinear corrections to the heavy-quark
production of gluon density at low $x$ at hot-spot point are
implemented. The heavy-quark structure functions can be improved
by comparison with both experimental data and parameterization
models. Above $Q^{2}{\geq}m^{2}_{t}$, the numerical values of the
top quark reduced cross section at the FCC-he collider become
large in the collinear approach, which needs further
investigation. For top quark pair production, which will be an
important production channel at both LHeC and FCC-eh, the linear
and nonlinear reduced cross sections were determined. The results
of the top reduced cross section are available for high
inelasticity, defined with respect to the center-of-mass energy.
With this method bounds $\sigma_{r}^{t\overline{t}}|_{y=1}$ are
derived in the collinear approach in the LO up to NNLO
approximations at FCC-he center-of-mass energies. Furthermore, the
importance of the longitudinal structure function measurements in
top quark production at future ep collider energies are studied,
investigating the ratio
$F_{L}^{t\overline{t}}/F_{2}^{t\overline{t}}$. Ratios of top quark
structure functions are studied in the LO up to NNLO
approximations taking into account the top quark mass in the
rescaling variable, which is important for the FCC-he.
Furthermore, this paper proposes to perform a study of the
$t\overline{t}$ production at the high inelasticity at the future
FCC-he collider.\\

\subsection{ACKNOWLEDGMENTS}
The author is thankful to the Razi University for financial
support of this project. The author is especially grateful to M.
Klein and C.Schwanenberger for carefully reading the paper.\\

%%%%%%%%%%%%%%%%%%%%%%%%%%%%%%%%%%%%%%%%%%%%%%%%%%%%%%%%%%%%%%
\subsection{APPENDIX A}
In the high energy regime, defined by $x{\ll}1$, the coefficient
functions have the compact forms [25]
\begin{eqnarray}
B_{k,g}^{(2)}(x,a)&=&\beta{\ln}(1/x)[R_{k,g}^{(2)}(1,a)+4C_{A}R_{k,g}^{(1)}(1,a)L_{\mu}\nonumber\\
&&+8C_{A}^{2}B_{k,g}^{(0)}(1,a)L^{2}_{\mu}],
\end{eqnarray}
with
\begin{eqnarray}
R_{2,g}^{(2)}(1,a)&=&
\frac{32}{27}C_{A}^{2}[46+(71-92a)J(a)\nonumber\\
&&+3(13-10a)I(a)-9(1-a)K(a)],\nonumber\\
R_{L,g}^{(2)}(1,a)&=&\frac{64}{27}C_{A}^{2}x_{2}\{34+240a-[3+136a+480a^{2}]J(a)\nonumber\\
&&+3[3+4a(1-6a)]I(a)+18a(1+3a)K(a)\},\nonumber\\
R_{2,g}^{(1)}(1,a)&=&
\frac{8}{9}C_{A}[5+(13-10a)J(a)\nonumber\\
&&+6(1-a)I(a)],\nonumber\\
R_{L,g}^{(1)}(1,a)&=&-\frac{16}{9}C_{A}x_{2}\{1-12a-[3+4a(1-6a)]J(a)\nonumber\\
&&+12a[1+3a]I(a)\},\nonumber\\
B_{2,g}^{(0)}(1,a)&=&
\frac{2}{3}[1+2(1-a)J(a),\nonumber\\
B_{L,g}^{(0)}(1,a)&=&\frac{4}{3}x_{2}\{1+6a-4a[1+3a]J(a),
\end{eqnarray}
where
\begin{eqnarray}
K(a)&=&-\sqrt{x_{2}}~[4(\zeta_{3}+\mathrm{Li}_{3}(-t)-\mathrm{Li}_{2}(-t){\ln}t\nonumber\\
&&-2S_{1,2}(-t))+2{\ln}(ax_{2})(\zeta_{2}+2\mathrm{Li}_{2}(-t))\nonumber\\
&&-\frac{1}{3}{\ln}^{3}t
-{\ln}^{2}(ax_{2}){\ln}t+{\ln}(ax_{2}){\ln}^{2}t],\nonumber\\
I(a)&=&-\sqrt{x_{2}}~
[\zeta_{2}+\frac{1}{2}{\ln}^{2}t-{\ln}(ax_{2}){\ln}t+2\mathrm{Li}_{2}(-t)],\nonumber\\
J(a)&=&-\sqrt{x_{2}}~ {\ln}t,\nonumber\\
t&=&\frac{1-\sqrt{x_{2}}}{1+\sqrt{x_{2}}},\nonumber\\
x_{2}&=&\frac{1}{1+4a},\nonumber\\
L_{\mu}&=&\ln{\frac{4m^{2}}{\mu^{2}}},\nonumber\\
\end{eqnarray}
where
\begin{eqnarray}
\mathrm{Li}_{2}(x)&=&-\int_{0}^{1}\frac{dy}{y}{\ln}(1-xy),\nonumber\\
\mathrm{Li}_{3}(x)&=&-\int_{0}^{1}\frac{dy}{y}{\ln}(y){\ln}(1-xy),\nonumber\\
S_{1,2}(x)&=&\frac{1}{2}\int_{0}^{1}\frac{dy}{y}{\ln}^{2}(1-xy),\nonumber\\
\end{eqnarray}
are the dilogarithmic function $\mathrm{Li}_{2}(x)$, the
trilogarithmic
function $\mathrm{Li}_{3}(x)$ and Nilsen Polylogarithm $S_{1,2}(x)$.\\
%%%%%%%%%%%%%%%%%%%%%%%%%%%%%%%%%%%%%%%%%%%%%%%%%%%%%%%%%%%%%
\subsection{References}
1. P.Agostini  et al. [LHeC Collaboration and FCC-he Study Group ], J. Phys. G: Nucl. Part. Phys. {\bf48}, 110501 (2021).\\
2. A. Abada et al., [FCC Collaboration], Eur.Phys.J.C {\bf79}, 474 (2019).\\
3. M.Klein, arXiv:1802.04317;  Annalen Phys. {\bf528},
138(2016).\\
4. R.A.Khalek et al., SciPost Phys.{\bf7}, 051(2019).\\
5. A. Caldwell and  M. Wing, Eur. Phys. J. C {\bf76}, 463 (2016);
A. Caldwell, et al., arXiv:1812.08110.\\
6. G. R. Boroun, ; Phys.Lett.B {\bf744}, 142
 (2015); Phys.Lett.B {\bf741}, 197 (2015); Physics of Particles and Nuclei Letters
{\bf15}, 387(2018); Chin.Phys. C {\bf41},  013104 (2017).\\
7. I. A. Sarmiento-Alvarado, A. O. Bouzas, and F. Larios, J.
Phys. G {\bf42}, 085001 (2015).\\
8. Turk Cakir, A. Yilmaz, H. Denizli, A. Senol, H. Karadeniz, and
O. Cakir, Adv. High Energy Phys.{\bf 2017}, 1572053 (2017).\\
9. H. Sun (LHeC/FCC-eh top physics
Study Group), PoS DIS2018, 186 (2018).\\
10. C. Schwanenberger,
PoS EPS-HEP2019, 635 (2020).\\
11. W. Liu and H. Sun, Phys. Rev. D {\bf100}, 015011 (2019); B.
Yang, B. Hou, H. Zhang, and N. Liu, Phys. Rev. D {\bf99},
095002 (2019); B. Rezaei and G. R. Boroun, EPL {\bf130}, 51002 (2020).\\
12. H.Khanpour, Nucl.Phys.B {\bf958}, 115141 (2020).\\
13. G.R.Boroun and B.Rezaei, EPL {\bf133}, 61002 (2021).\\
14. M.Gao and J.Gao, Phys. Rev. D {\bf104}, 053005 (2021).\\
15. V.A.Okorokov, J. Phys.: Conf. Ser. 1690, 012006, (2020).\\
16. O.B.Bylund, arXiv:2103.14772.\\
17. E.R.Nocera, M.Ubiali and C.Voisey, JHEP {\bf05}, 067 (2020).\\
18. P.A.Zyla, et al. (Particle Data Group) (2020) Review of
Particle Physics. Progress of Theoretical and Experimental
Physics, 2020, 083C01.\\
19. CMS Collaboration, Eur.Phys.J.C {\bf79}, 368 (2019); Phys. Rev. D {\bf93}, 072004 (2016);
ATLAS Collaboration, Phys. Lett. B {\bf810},  135797(2020); Eur. Phys. J. C {\bf79}, 290 (2019).\\
20. Tie-Jiun Hou et al., Phys.Rev.D{\bf103}, 014013(2021).\\
21. R.S.Thorne, arXiv:hep-ph/9805298(1998).\\
22. A.D.Martin W.J.Stirling and R.S.Thorne, Phys.Lett.B {\bf 636}, 259(2006).\\
23. J.Bl$\ddot{\mathrm{u}}$mlein, A.De Freitas, C.Schneider and
K.Sch$\ddot{\mathrm{o}}$nwald, Phys. Lett.B {\bf782},
 362(2018).\\
24.  S.Alekhin, J. Bl$\ddot{\mathrm{u}}$mlein and S. Moch, Phys. Rev. D {\bf102}, 054014 (2020).\\
25. A.V.Kotikov, A.V.Lipatov and P.Zhang, Phys. Rev. D {\bf104}, 054042 (2021).\\
26. M.A. Kimber, A.D. Martin, M.G. Ryskin, Phys. Rev. D {\bf63},
114027 (2001); G. Watt, A.D. Martin, M.G. Ryskin, Eur. Phys. J. C
{\bf31}, 73 (2003).\\
27. A.V. Kotikov, G. Parente, Nucl. Phys. B {\bf549}, 242 (1999);
A.Yu. Illarionov, A.V. Kotikov, G. Parente, Phys. Part. Nucl. 39,
307 (2008); L. Mankiewicz, A. Saalfeld, T. Weigl, Phys. Lett. B
 {\bf393}, 175 (1997).\\
28. M.M.Block and L.Durand, arXiv: 0902.0372 [hep-ph](2009); E.L.
Berger, M.M. Block and Chung-I Tan, Phys.Rev.Lett. {\bf98}, 242001
(2007); M.M. Block, L. Durand and D.W. McKay, Phys.Rev.D {\bf79},
014031 (2009).\\
29. M.A.G.Aivazis, J.C.Collins, F.I.Olness and W.-K.Tung,
Phys.Rev.D {\bf50}, 3102 (1994).\\
30. A.Y.Illarionov, B.A.Kniehl and A.V.Kotikov,
 Phys.Lett.B {\bf 663}, 66 (2008); A.Y.Illarionov and A.V.Kotikov,
 Phys.Atom.Nucl {\bf 75}, 1234 (2012).\\
31. G.R.Boroun and B.Rezaei, Int.J.Mod.Phys.E {\bf24},
1550063(2015);  Nucl.Phys.A {\bf929}, 119(2014); EPL {\bf100}, 41001(2012); J.Exp.Theor.Phys. {\bf115}, 427(2012); Nucl.Phys.B {\bf857}, 143(2012).\\
32. J.Lan et al., Phys. Rev. D {\bf102}, 014020 (2020);
N.N.Nikolaev and V.R.Zoller, Phys.Atom.Nucl. \textbf{73},
672(2010); A.~V.~Kotikov, A.~V.~Lipatov, G.~Parente and
N.~P.~Zotov, Eur.
Phys. J. C {\bf 26}, 51 (2002).\\
33. G.R.Boroun, Nucl.Phys.B {\bf884}, 684(2014).\\
34. D.B.Clark, E.Godat and  F.I.Olness, Comput.Phys.Commun.
{\bf216}, 126 (2017).\\
35. A.V.Giannini and F.O.Dur$\widetilde{a}$es, Phys.Rev.D {\bf88}, 114004(2013).\\
36. G.R.Boroun and B.Rezaei, Phys. Rev. C {\bf103}, 065202 (2021).\\
37. R.Wang and X.Chen, Chinese Phys.C {\bf41}, 053103(2017).\\
38. G.R.Boroun and S.Zarrin, Eur.Phys.J.Plus {\bf128}, 119(2013).\\
39. B.Rezaei and G.R.Boroun, Phys.Lett.B {\bf692}, 247(2010).\\
40. D. Britzger, C. Ewerz, S. Glazov, O. Nachtmann, and S.
Schmitt, Phys. Rev. D {\bf100}, 114007 (2019).\\
41. G.R.Boroun, Eur.Phys.J.A {\bf42}, 251(2009); Eur.Phys.J.A {\bf43}, 335(2010).\\
42. L. V. Gribov, E. M. Levin and M. G. Ryskin, Phys. Rep.
{\bf100}, 1 (1983).\\
43. A. H. Mueller and Jianwei Qiu, Nucl. Phys. B {\bf268}, 427
(1986).\\
44. G.R.Boroun and B.Rezaei, Eur.Phys.J.C {\bf81}, 851 (2021);
A.V. Kotikov, JETP Lett. {\bf111}, 67 (2020).\\
45. J.Kwiecinski et al., Phys.Rev.D {\bf42}, 3645 (1990).\\
46. PARTICLE DATA GROUP collaboration, ${Review~ of~
particle~ physics}$, Phys.Rev.D {\bf98}, 030001 (2018).\\
47. LHC Higgs Cross Section Working Group collaboration, arXiv:1610.07922.\\
48. NNPDF Collaboration (Ball R. D. et al.), Eur. Phys. J. C
{\bf77}, 663 (2017).\\
49. G. R. Boroun, Chin. Phys. C {\bf45}, 063105 (2021).\\

%%%%%%%%%%%%%%%%%%%%%%%%%%%%%%%%%%%%%%%%%%%%%%%%%%%%%%%%%%
%\begin{figure} \centering
%\includegraphics[width=0.45\textwidth]{Fig2new.eps}
%\caption{The ratio
%$F_{2}^{t\overline{t}}/\sigma_{r}^{t\overline{t}}$ are shown
%importance of top quark longitudinal structure function
%measurements at  $Q^{2}{\geq}m^{2}_{t}$ for
%$\sqrt{s}=3.5~\mathrm{TeV}$. The predictions obtained with
%analytical parameterization of the gluon density in a proton at LO
%and collinear approach at LO up to NNLO approximations. }
%\label{Fig2}
%\end{figure}
%\begin{figure} \centering
%\includegraphics[width=0.45\textwidth]{Fig5new.eps}
%\caption{The same as Fig.2 for $\sqrt{s}=1.3~\mathrm{TeV}$ at
%$Q^{2}{\leq}m^{2}_{t}$. } \label{Fig5}
%\end{figure}
%\begin{figure} \centering
%\includegraphics[width=0.55\textwidth]{Fig6new.eps}
%\caption{The same as Fig.3 for $\sqrt{s}=1.3~\mathrm{TeV}$ at
%$Q^{2}{\leq}m^{2}_{t}$.} \label{Fig6}
%\end{figure}
%\begin{figure} \centering
%\includegraphics[width=0.55\textwidth]{Fig9new.eps}
%\caption{The same as Fig.8 for the ratio
%$F_{2}^{t\overline{t}}/\sigma_{r}^{t\overline{t}}$.} \label{Fig1}
%\end{figure}
%\begin{figure} \centering
%\includegraphics[width=0.55\textwidth]{Fig10new.eps}
%\caption{The same as Fig.8 for the ratio
%$R^{t\overline{t}}=\frac{F_{L}^{t\overline{t}}}{F_{2}^{t\overline{t}}}$.}
%\label{Fig1}
%\end{figure}
%%%%%%%%%%%%%%%%%%%%%%%%%%%%%%%%%%%%%%%%%%%%%%%%%%%%%%%%%%
\end{document}